\documentclass[12pt]{article}
\usepackage{amsmath,amssymb}
\usepackage{hyperref}
\usepackage{units}
\usepackage{color}
\usepackage[T1]{fontenc}
\usepackage[utf8]{inputenc}
\usepackage{authblk}
\usepackage{bm,latexsym,mathrsfs,enumerate}
\renewcommand{\vec}[1]{\bm{#1}}

\title{Explicit matrix representation for the Hamiltonian of the one dimensional spin $1/2$ Ising model in mutually orthogonal external magnetic fields.\thanks{%
PACS: 75.10.Pq, 75.10.Jm, 64.60.Cn}}
\author[]{Kunle Adegoke\thanks{Corresponding author: adegoke00@gmail.com}}
\author[]{Henry Otobrise}
\author[]{Tolulope Famoroti}
\author[]{\mbox{Adenike Olatinwo}}
\author[]{Afees Tiamiyu}
\author[]{Funmi Akintujoye}
\affil{Department of Physics and Engineering Physics, \mbox{Obafemi Awolowo University}, Ile-Ife, 220005 Nigeria}

\begin{document}

\date{}

\maketitle

\begin{abstract}
\noindent We derive an explicit matrix representation for the Hamiltonian of the Ising model in mutually orthogonal external magnetic fields, using as basis the eigenstates of a system of non-interacting spin~$1/2$ particles in external magnetic fields. We subsequently apply our results to obtain an analytical expression for the ground state energy per spin, to the fourth order in the exchange integral, for the Ising model in perpendicular external fields.
\end{abstract}
\pagebreak
\tableofcontents

\section{Introduction}

Field-induced effects in low-dimensional quantum spin systems have been studied for a long time~\cite{affleck99, langari06}. Hamiltonian models incorporating external magnetic fields are gaining popularity among experimentalists as well as theoreticians (see references~\cite{sen,kenzelmann02,ovchin,dmitriev04}). A longitudinal field is often introduced mainly to facilitate the calculation of order parameter and associated susceptibility as can be seen for example in references~\cite{marland81,barber,hamer}, and a transverse field to introduce quantum fluctuations~\cite{pfeuty,rieger}.

\bigskip

Our main objective in this paper is to give an explicit matrix representation for the Hamiltonian of a system of $N$ \mbox{spin-$1/2$} particles on a cyclic one dimensional lattice chain, interacting via nearest neighbour exchange, in the presence of transverse and longitudinal external magnetic fields.

\bigskip

The Hamiltonian, $H$, is
\begin{equation}\label{equ.ir12ul8}
H =  - h_x \sum_{i = 1}^N {S_i^x }  - h_y \sum_{i = 1}^N {S_i^y }  - h_z \sum_{i = 1}^N {S_i^z }  - J\sum_{i = 1}^N {S_i^z S_{i + 1}^z }\,,
\end{equation} where $h_x$ and $h_y$ are the uniform external transverse magnetic fields, $h_z$ is the uniform longitudinal field, $J$ is the nearest neighbour exchange interaction, $S_i$ are the usual spin-$1/2$ operators and the fields $h_x$, $h_y$ and $h_z$ are measured in units where the splitting factor and Bohr magneton are equal to unity. Periodic boundary condition is assumed so that \mbox{$S_{N+i}^z\equiv S_i^z$}, and so on. The parameters $h_x$, $h_y$, $h_z$ and $J$ are all assumed to be \mbox{non-negative}.

\bigskip

It is convenient to write $H = H_F  + H_I$, where
\[
\begin{split}
&\quad H_I  =  - J\sum_{i = 1}^N {S_i^z S_{i + 1}^z }\\
&\mbox{and}\\
&\quad H_F =  - h_x \sum_{i = 1}^N {S_i^x }  - h_y \sum_{i = 1}^N {S_i^y }  - h_z \sum_{i = 1}^N {S_i^z }\,.
\end{split}
\]

$H_F$ describes a system of $N$ non-interacting spin~$1/2$ particles in mutually orthogonal external magnetic fields.

\bigskip

The model~\eqref{equ.ir12ul8} has been widely studied for various combinations of the parameters $h_x$, $h_y$, $h_z$ and $J$, especially for phase transitions (see~\cite{sen,ovchin,duttaetal2012} and the references therein). Our aim is to give an explicit matrix representation for $H$, using the eigenstates of $H_F$ as basis.

\bigskip

Throughout this paper we will make use of the following identities which hold for \mbox{$j,k\in\{0,1\}$}:
\begin{equation}\label{equ.cgfdre}
\begin{split}
&j\equiv\sin^2(j\pi/2),\quad 1-j\equiv\cos^2(j\pi/2)\,,\\
&\\
&\delta _{jk}  \equiv 1 - j - k + 2jk \equiv\cos^2 \left\{ {(j - k){\pi  \mathord{\left/
 {\vphantom {\pi  2}} \right.
 \kern-\nulldelimiterspace} 2}} \right\}\,,\\
&\\
&j + k - 2jk\equiv\sin^2 \left\{ {(j - k){\pi  \mathord{\left/
 {\vphantom {\pi  2}} \right.
 \kern-\nulldelimiterspace} 2}} \right\}\,,\\
&\\
&( - 1)^j \delta _{jk}  \equiv 1 - j - k \equiv \delta _{jk}  - 2jk \equiv \cos \left\{ {(j + k){\pi  \mathord{\left/
 {\vphantom {\pi  2}} \right.
 \kern-\nulldelimiterspace} 2}} \right\}\,,\\
&\\
&\mbox{in particular }( - 1)^j  \equiv 1 - 2j\equiv\cos j\pi,\quad( - 1)^{j - 1}  \equiv 2j - 1\,,\\
&\\
&( - 1)^j  + ( - 1)^k  \equiv 2( - 1)^j \delta _{jk} ,\quad( - 1)^{j + k}  \equiv 2\delta _{jk}  - 1\equiv \cos \{(j-k)\pi\},\\
&\\
&j\delta _{jk}  \equiv jk\,.
\end{split}
\end{equation}

\section{Quantization of a system of non-interacting spin~$1/2$ particles in external magnetic fields}

A system of $N$ non-interacting spin~$1/2$ particles in mutually orthogonal external magnetic fields $h_x$, $h_y$ and $h_z$ is described by the Kronecker sum Hamiltonian
\[
H_F  = H_{F_1 }  \oplus H_{F_2 }  \oplus  \cdots  \oplus H_{F_N } 
\]
where, for $j,k\in\{0,1\}$, each single particle Hamiltonian $H_{F_i}$, at the $ith$ site, has the matrix elements, in unit of $\hbar$,
\[
\begin{split}
\left\langle {\lambda _j } \right|H_{F_i } \left| {\lambda _k } \right\rangle  &=  - \frac{{h_z }}{2}\cos j\pi \cos ^2 \left\{ {\left( {j - k} \right)\frac{\pi }{2}} \right\}\\
&\\
&\qquad- \left[ {\frac{a}{2}\cos ^2 \left( {\frac{{j\pi }}{2}} \right) + \frac{{a^ *  }}{2}\cos ^2 \left( {\frac{{k\pi }}{2}} \right)} \right]\sin ^2 \left\{ {\left( {j - k} \right)\frac{\pi }{2}} \right\}
\end{split}
\]

with respect to the eigenstates $\{\left| {\lambda _0 } \right\rangle, \left| {\lambda _1 } \right\rangle\}$ of the spin~$1/2$ operator $S_i^z$, whose elements, in unit of $\hbar$, are 
\[
\left\langle {\lambda _j } \right|S_i^z \left| {\lambda _k } \right\rangle  = \frac{{\cos j\pi }}{2}\cos ^2 \left\{ {\left( {j - k} \right)\frac{\pi }{2}} \right\} = \lambda _j \cos ^2 \left\{ {\left( {j - k} \right)\frac{\pi }{2}} \right\}\,.
\]
The remaining two spin~$1/2$ operators $S_i^x$ and $S_i^y$ have matrix elements given by 
\[
\begin{split}
\left\langle {\lambda _j } \right|S_i^x \left| {\lambda _k } \right\rangle  &= \frac{1}{2}\sin ^2 \left\{ {\left( {j - k} \right)\frac{\pi }{2}} \right\}\\
&\mbox{and}\\
\left\langle {\lambda _j } \right|S_i^y \left| {\lambda _k } \right\rangle  &= \frac{{ - i\cos j\pi }}{2}\sin ^2 \left\{ {\left( {j - k} \right)\frac{\pi }{2}} \right\}\,.
\end{split}
\]

\bigskip

Parameters $h_x$, $h_y$ and $h_z$ are the external magnetic fields and $a=h_x-ih_y$. 

\bigskip

Explicitly,
\[
\begin{split}
H_{F_i }  &=- h_x S_i^x  - h_y S_i^y  - h_z S_i^z\\
&\\
&=- \frac{1}{2}\left( {\begin{array}{*{20}c}
   {h_z } & {h_x  - ih_y }  \\
   {h_x  + ih_y } & { - h_z }  \\
\end{array}} \right)\,.
\end{split}
\]

\subsection{Change of basis via the eigenstates of the single particle Hamiltonian}
Solving the eigenvalue equation \mbox{$H_{F_i } \left| {\varepsilon _j } \right\rangle  = \varepsilon _j \left| {\varepsilon _j } \right\rangle $}, the normalized eigenstates $\left| {\varepsilon _j } \right\rangle$, $j\in\{0,1\}$, are found to be
\[
\left| {\varepsilon _j } \right\rangle  =   ac_j\left| {\lambda _0 } \right\rangle  + b_jc_j \left| {\lambda _1 } \right\rangle  \,,
\]
with corresponding eigenvalues
\begin{equation}\label{equ.qgo2lrw}
\varepsilon_j=-h/2\cos j\pi\,,
\end{equation}
where
\begin{equation}
\begin{split}
&h = \left( {h_x^2  + h_y^2  + h_z^2 } \right)^{1/2}\,,\\
&\\
&a = h_x  - ih_y ,\quad b_j  = -h\cos j\pi - h_z\\
&\\
&\mbox{and}\\
&\\
&c_j  = -\frac{{\cos j\pi}}{{(2h)^{1/2} \left( {h +h_z\cos j\pi } \right)^{1/2} }} = \frac{{\left( {h +h_z\cos j\pi } \right)^{1/2} }}{{(2h)^{1/2} b_j }}\,.
\end{split}
\end{equation}
Note that
\[
\begin{split}
&a^ *  a =  - b_0 b_1  = h_x^2  + h_y^2  = h^2  - h_z^2\,,\\
&\\
&a^ *  + a = 2h_x ,\quad a^ *   - a = 2ih_y\,,\\
&\\
&b_j b_k  = \left( {h + h_z\cos j\pi } \right)^2 \cos ^2 \left\{ {\left( {j - k} \right)\frac{\pi }{2}} \right\}  - \left( {h^2  - h_z^2 } \right)\sin ^2 \left\{ {\left( {j - k} \right)\frac{\pi }{2}} \right\}\\
&\\
&c_j c_k  = \frac{1}{{2h}}\left( {\frac{{\cos ^2 \left\{ {{{\left( {j - k} \right)\pi } \mathord{\left/
 {\vphantom {{\left( {j - k} \right)\pi } 2}} \right.
 \kern-\nulldelimiterspace} 2}} \right\}}}{{h + h_z \cos j\pi }} - \frac{{\sin ^2 \left\{ {{{\left( {j - k} \right)\pi } \mathord{\left/
 {\vphantom {{\left( {j - k} \right)\pi } 2}} \right.
 \kern-\nulldelimiterspace} 2}} \right\}}}{{\left( {h^2  - h_z^2 } \right)^{1/2} }}} \right)\,,\\
&\\
&\mbox{and}\\
&\\
&a^ *  a\sum_{j = 0}^1 {c_j^2 }  = 1,\quad\sum_{j = 0}^1 {c_j^2 b_j }  = 0\,.
\end{split}
\]
The diagonalizing matrix $P$ has elements $P_{jk}=ac_k\cos^2{(j\pi/2)}+jb_kc_k$, for $j,k\in\{0,1\}$. Thus, $H_{F_i}$ is similar to the diagonal matrix $D$ having elements $D_{jk}=\varepsilon_j\cos^2{({(j-k)\pi/2})}$, that is
\[
\begin{split}
&H_{F_i }  = PDP^ \dagger\,,\\
&\\
&P = \left( {\begin{array}{*{20}c}
   {c_0 a} & {c_1 a}  \\
   {c_0 b_0 } & {c_1 b_1 }  \\
\end{array}} \right),\quad D = \left( {\begin{array}{*{20}c}
   {\varepsilon _0 } & 0  \\
   0 & {\varepsilon _1 }  \\
\end{array}} \right)\,.
\end{split}
\]
With respect to the new basis, $\left\{ {\left| {\varepsilon _0 } \right\rangle ,\left| {\varepsilon _1 } \right\rangle } \right\}$, and for $j,k\in\{0,1\}$, the Pauli spin matrices have the representation 
\[
\begin{split}
\left\langle {\varepsilon _j } \right|S_i^x \left| {\varepsilon _k } \right\rangle  &=  - \frac{{h_x }}{{2h}}\cos j\pi \cos ^2 \left\{ {\left( {j - k} \right)\frac{\pi }{2}} \right\}\\
&\\
&\qquad+ \left[ {\left( {h\cos j\pi  + h_z } \right)a + \left( {h\cos k\pi  + h_z } \right)a^ *  } \right]\frac{{\sin ^2 \left\{ {{{\left( {j - k} \right)\pi } \mathord{\left/
 {\vphantom {{\left( {j - k} \right)\pi } 2}} \right.
 \kern-\nulldelimiterspace} 2}} \right\}}}{{4h\left( {h^2  - h_z^2 } \right)^{1/2} }}\,,
\end{split}
\]
\[
\begin{split}
\left\langle {\varepsilon _j } \right|S_i^y \left| {\varepsilon _k } \right\rangle  &=  - \frac{{h_y }}{{2h}}\cos j\pi \cos ^2 \left\{ {\left( {j - k} \right)\frac{\pi }{2}} \right\}\\
&\\
&\qquad+ \left[ {\left( {h\cos j\pi  + h_z } \right)a - \left( {h\cos k\pi  + h_z } \right)a^ *  } \right]\frac{{i\sin ^2 \left\{ {{{\left( {j - k} \right)\pi } \mathord{\left/
 {\vphantom {{\left( {j - k} \right)\pi } 2}} \right.
 \kern-\nulldelimiterspace} 2}} \right\}}}{{4h\left( {h^2  - h_z^2 } \right)^{1/2} }}
\end{split}
\]
and
\[
\begin{split}
\left\langle {\varepsilon _j } \right|S_i^z \left| {\varepsilon _k } \right\rangle  &=  - \frac{{h_z }}{{2h}}\cos j\pi \cos ^2 \left\{ {\left( {j - k} \right)\frac{\pi }{2}} \right\}\\
&\\
&\qquad- \frac{{\left( {h^2  - h_z^2 } \right)^{1/2} }}{{2h}}\sin ^2 \left\{ {\left( {j - k} \right)\frac{\pi }{2}} \right\}\,.
\end{split}
\]

\subsection{General basis states for the matrix representation of one dimensional spin~$1/2$ Hamiltonian systems}

Since $H_F$ is a Hermitian operator that lives in a $2^N$ dimensional Hilbert space, $\mathcal{H}$, its eigenstates form a complete orthonormal basis, suitable for giving matrix representations for operators living in $\mathcal{H}$ and with the same conditions at the boundary. The eigenvalue equation for $H_F$ is
\[
H_F \left| {E_r } \right\rangle  = E_r \left| {E_r } \right\rangle,\quad r=0,1,2,\ldots,2^N-1\,.
\]

For each $r$ the eigenstate $\left| {E_r } \right\rangle$ is a direct product of the eigenstates of $H_{F_i}$ while the eigenvalue is the sum of the respective eigenvalues $\varepsilon_i$, that is
\[
\begin{split}
\left| {E_r } \right\rangle  &= \left| {\varepsilon _{r_1 } } \right\rangle  \otimes \left| {\varepsilon _{r_2 } } \right\rangle  \otimes  \cdots  \otimes \left| {\varepsilon _{r_N } } \right\rangle  = \prod_{i = 1}^N {\left| {\varepsilon _{r_i } } \right\rangle }\\
&\\
&\mbox{and}\\
&\\
E_r  &= \varepsilon _{r_1 }  + \varepsilon _{r_2 }  +  \cdots  + \varepsilon _{r_N }  = \sum_{i = 1}^N {\varepsilon _{r_i } }\,,
\end{split}
\]
where
\[
r_i  = \sin ^2 \left\{ {\left( {\left\lfloor {\frac{r}{{2^{N - i} }}} \right\rfloor } \right)\frac{\pi }{2}} \right\},\quad i=1,2,\ldots N\,,
\]
where $\left\lfloor z \right\rfloor$, the {\em floor} of $z$, is the smallest integer not greater than $z$. Thus each state $\left| {E_r } \right\rangle$ is uniquely represented by a binary vector \mbox{$\vec r=\left(r_1,r_2,\ldots,r_N\right)$}.

\bigskip

Thus, any operator $A$ in $\mathcal{H}$ has the matrix representation $A$ with elements given by
\[
A_{rs}  = \left\langle {E_r } \right|A\left| {E_s } \right\rangle\,. 
\]
Using~\eqref{equ.qgo2lrw} we get
\begin{equation}\label{equ.cvuwlu9}
E_r  = h\sum_{i = 1}^N {r_i }  - \frac{{Nh}}{2} = hm_r - \frac{{Nh}}{2}\,.
\end{equation}
Note that $m_r=\sum_{i=1}^Nr_i$ counts the number of $\left| {\varepsilon_1 } \right\rangle$ states in the direct product state $\left| {E_r } \right\rangle$. The degeneracy of the state $\left| {E_r } \right\rangle$ is therefore \mbox{$g(E_r)={}^NC_{m_r} $}. Thus only the ground state and the most excited state are non-degenerate.

\section{Quantization of the one dimensional spin~$1/2$ Ising model in external magnetic fields}

\subsection*{Explicit matrix representation}
Since $H_F$ is diagonal in the basis $\{\left| {E_r } \right\rangle\}$, the only task is to find the matrix elements of $H_I$ and then add them to those of $H_F$. We have
\begin{equation}\label{equ.plwwsaf}
\begin{split}
H_{I_{rs} }  = \left\langle {E_r } \right|H_I \left| {E_s } \right\rangle  &=  - J\sum_{i = 1}^N {\left\langle {E_r } \right|S_i^z S_{i + 1}^z \left| {E_s } \right\rangle }\\
&=  - J\sum_{i = 1}^N {d_{i_{rs} } S_{i_{r_i s_i } }^z S_{i + 1_{r_{i + 1} s_{i + 1} } }^z }\,,
\end{split}
\end{equation}
where $S_{k_{r_k s_k } }^z  = \left\langle {\varepsilon _{r_k } } \right|S_k^z \left| {\varepsilon _{s_k } } \right\rangle $ and where we have introduced an \mbox{$N-$dimensional} vector $\vec d$ whose components are $2^N\times 2^N$ symmetric binary matrices $d_i$ defined by
\begin{equation}\label{equ.o420h6u}
d_{i_{rs} }  = \prod_{\scriptstyle j = 1 \hfill \atop 
  {\scriptstyle j \ne i \hfill \atop 
  \scriptstyle j \ne i + 1 \hfill}}^N {\delta _{r_i s_i } }\,.
\end{equation}
Thus $ d_{i_{rs} }=1 $ if either the two vectors $\vec r$ and $\vec s$ are one and the same vector, that is $\vec r=\vec s$, or they differ only at the consecutive $i^{th}$ and $(i+1)^{th}$ entries, otherwise $ d_{i_{rs} } = 0$.

\bigskip

Note that
\begin{equation}\label{equ.h54j74c}
\delta _{r_i s_i } \delta _{r_{i + 1} s_{i + 1} } d_{i_{rs} }  = \delta _{r_i s_i } c_{i_{rs} }  = \delta _{rs}\,,
\end{equation}
where we have introduced another \mbox{$N-$dimensional} vector $\vec c$ whose components are $2^N\times 2^N$ symmetric binary matrices $c_i$ with elements given by
\begin{equation}\label{equ.zjae2nl}
c_{i_{rs} }  = \prod_{\scriptstyle j = 1 \hfill \atop 
  \scriptstyle j \ne i \hfill}^N {\delta _{r_i s_i } }\,.
\end{equation}
Thus $ c_{i_{rs} }=1 $ if either the two vectors $\vec r$ and $\vec s$ are one and the same vector, $\vec r=\vec s$, or they differ only at the $i^{th}$ component, otherwise $ c_{i_{rs} } = 0$.

\bigskip
 
Motivated by the definitions in \eqref{equ.o420h6u}, \eqref{equ.h54j74c} and \eqref{equ.zjae2nl} we introduce two more \mbox{$N-$dimensional} vectors, $\vec\alpha$ and $\vec\beta$, whose components are $2^N\times 2^N$ symmetric binary matrices, in terms of which the $c_i$ and $d_i$ matrices may also be expressed. The $\alpha_i$ and $\beta_i$ matrices are defined through their elements by
\[
\begin{split}
\alpha _{i_{rs} }  &= \delta _{r_i s_i }  = \cos ^2 \left\{ {\left( {r_i  - s_i } \right){\pi  \mathord{\left/
 {\vphantom {\pi  2}} \right.
 \kern-\nulldelimiterspace} 2}} \right\},\\
&\\
\beta _{i_{rs} }  &= \delta _{r_i s_i } \delta _{r_{i + 1} s_{i + 1} }\\
&\\
& = \alpha _{i_{rs} } \alpha _{i + 1_{rs} }  = \cos ^2 \left\{ {\left( {r_i  - s_i } \right){\pi  \mathord{\left/
 {\vphantom {\pi  2}} \right.
 \kern-\nulldelimiterspace} 2}} \right\}\cos ^2 \left\{ {\left( {r_{i + 1}  - s_{i + 1} } \right){\pi  \mathord{\left/
 {\vphantom {\pi  2}} \right.
 \kern-\nulldelimiterspace} 2}} \right\}\,.
\end{split}
\]
It is straightforward to verify the following properties for the $\alpha_i$ and $\beta_i$ matrices:
\begin{equation}\label{equ.zzwhk4b}
\begin{split}
&\alpha _i \alpha _j  =\alpha _j \alpha _i= 2^{N - 1} \delta _{ij} \alpha _i  + 2^{N - 2} (1 - \delta _{ij} )J_{2^N } ,\\
&\\
&\beta _i \beta _j  =\beta _j \beta _i= 2^{N - 2} \delta _{ij} \beta _i  + (1 - \delta _{ij} )\left\{ {2^{N - 3} \alpha _{j} \delta _{j,i + 1}  + \left( {1 - \delta _{j,i + 1} } \right)2^{N - 4} J_{2^N } } \right\}\\
&\\
&\mbox{and}\\
&\\
&\alpha _i \beta _j  =\beta _j \alpha _i= 2^{N - 2} \delta _{ij} \alpha _i  + (1 - \delta _{ij} )\left\{ {2^{N - 2} \alpha _i \delta _{i,j + 1}  + \left( {1 - \delta _{i,j + 1} } \right)2^{N - 3} J_{2^N } } \right\},
\end{split}
\end{equation}
where
\[
J_{2^N }  = \left( {\begin{array}{*{20}c}
   1 & 1 &  \vdots  & 1  \\
   1 & 1 &  \vdots  & 1  \\
    \vdots  &  \vdots  &  \vdots  &  \vdots   \\
   1 & 1 &  \vdots  & 1  \\
\end{array}} \right)
\]
is the $2^N\times 2^N$ {\it all-ones} matrix. The $\alpha_i$ and $\beta_i$ matrices are singular and have trace equal to $2^N$. The eigenvalues of $\alpha_i$ are $2^{N-1}$ repeated twice and $0$ repeated $2^N-2$ times while those of $\beta_i$ are $2^{N-2}$ repeated four times and $0$ repeated $2^N-4$ times. Finally using multinomial expansion theorem and \eqref{equ.zzwhk4b}, it is readily established that the matrices $\alpha=\sum_{i=1}^N\alpha_i$ and $\beta=\sum_{i=1}^N\beta_i$ satisfy
\[
\begin{split}
\alpha^2  &= 2^{N - 1} \alpha  + 2^{N - 2} N(N - 1)J_{2^N } ,\\
&\\
\beta^2  &= 2^{N - 2} (\alpha  + \beta ) + 2^{N - 4} N(N - 3)J_{2^N }\\
&\\
&\mbox{and}\\
&\\
\alpha \beta  &= 2^{N - 1} \alpha  + 2^{N - 3} N(N - 2)J_{2^N }\,.
\end{split}
\]
It is now obvious that
\begin{equation}\label{equ.b189fh0}
\begin{split}
c_{i_{rs} }  &= \delta _{rs}  + (1 - \alpha _{i_{rs} } )\,\delta _{\alpha _{rs} ,N - 1}\\
&\\
&= \delta _{rs}  + (1 - \alpha _{i_{rs} } )\,\delta _{\beta _{rs} ,N - 2}\\
&\\
&= \delta _{rs}  + \delta _{\beta _{rs} ,N - 2} \,\cos ^2 \left( {\alpha _{i_{rs} }{\pi  \mathord{\left/
 {\vphantom {\pi  2}} \right.
 \kern-\nulldelimiterspace} 2}} \right)\,,
\end{split}
\end{equation}
\begin{equation}\label{equ.ncqrix2}
\begin{split}
d_{i_{rs} }  &= \delta _{rs}  + (1 - \alpha _{i_{rs} } )\alpha _{i + 1_{rs} } \delta _{\beta _{rs} ,N - 2}\\ 
&\qquad + (1 - \alpha _{i + 1_{rs} } )\alpha _{i_{rs} } \delta _{\beta _{rs} ,N - 2}\\ 
&\qquad\qquad + (1 - \alpha _{i_{rs} } )(1 - \alpha _{i + 1_{rs} } )\delta _{\beta _{rs} ,N - 3}\\
&\\ 
& = \delta _{rs}  + \delta _{\beta _{rs} ,N - 3}  + (\delta _{\beta _{rs} ,N - 2}  - \delta _{\beta _{rs} ,N - 3} )(\alpha _{i_{rs} }  + \alpha _{i + 1_{rs} } )\\
&\qquad\qquad + (\delta _{\beta _{rs} ,N - 3}  - 2\delta _{\beta _{rs} ,N - 2} )\alpha _{i_{rs} } \alpha _{i + 1_{rs} }\,.
\end{split}
\end{equation}
From \eqref{equ.b189fh0} and \eqref{equ.ncqrix2} we find
\[
c_{rs}=\sum_{i = 1}^N {c_{i_{rs} } }  = N\delta _{rs}  + \delta _{\beta _{rs} ,N - 2}
\]
and
\[
d_{rs}=\sum_{i = 1}^N {d_{i_{rs} } }  = N\delta _{rs}  + 2\delta _{\beta _{rs} ,N - 2}  + \delta _{\beta _{rs} ,N - 3}\,.
\]
Explicitly
\[
c_{i_{rs} }  = \left\{ {\begin{array}{*{20}c}
   \cos ^2 \left( {\alpha _{i_{rs} }{\pi  \mathord{\left/
 {\vphantom {\pi  2}} \right.
 \kern-\nulldelimiterspace} 2}} \right) & \mbox{ if }{\beta _{rs}  = N - 2}  \\
 &\\

   0 & \mbox{ if }{\beta _{rs}  < N - 2}  \\
   &\\

   1 & \mbox{ if }{r = s}\,,  \\
\end{array}} \right.
\]

\bigskip

\bigskip

\[
d_{i_{rs} }  = \left\{ {\begin{array}{*{20}c}
   \cos ^2 \left( {\alpha _{i_{rs} }{\pi  \mathord{\left/
 {\vphantom {\pi  2}} \right.
 \kern-\nulldelimiterspace} 2}} \right)\cos ^2 \left( {\alpha _{i+1_{rs} }{\pi  \mathord{\left/
 {\vphantom {\pi  2}} \right.
 \kern-\nulldelimiterspace} 2}} \right) & \mbox{ if }{\beta _{rs}  = N - 3}  \\
&\\
   \sin ^2 \left\{ {(\alpha _{i_{rs} }  - \alpha _{i+1_{rs} } ){\pi  \mathord{\left/
 {\vphantom {\pi  2}} \right.
 \kern-\nulldelimiterspace} 2}} \right\} & \mbox{ if }{\beta _{rs}  = N - 2}  \\
&\\
0 & \mbox{ if }{\beta _{rs}  < N - 2}  \\
&\\

   1 & {r = s}  \,,\\
\end{array}} \right.
\]

\bigskip

\bigskip

\[
c_{rs}  = \left\{ {\begin{array}{*{20}c}
   0 & \mbox{ if }{\beta _{rs}  < N - 2}  \\
   &\\

   1 & \mbox{ if }{\beta _{rs}  = N - 2}  \\
   &\\

   N & \mbox{ if }{r = s}  \\
\end{array}} \right.
\]

\bigskip

and

\bigskip

\[
d_{rs}  = \left\{ {\begin{array}{*{20}c}
   0 & \mbox{ if }{\beta _{rs}  < N - 3}  \\
   &\\

   1 & \mbox{ if }{\beta _{rs}  = N - 3}  \\
   &\\

   2 & \mbox{ if }{\beta _{rs}  = N - 2}  \\
   &\\

   N & \mbox{ if }{r = s}  \\
\end{array}} \right.\,.
\]
From the definitions of the $c_i$ and $d_i$ matrices the following additional properties are evident:
\begin{enumerate}
\item $c_i^n=2^{n-1}c_i$, $d_i^n=4^{n-1}d_i$, for $n\in\mathbb{Z^+}$.
\item The eigenvalues of $c_i$ are $0$ and $2$, each repeated $2^{N-1}$ times while those of $d_i$ are $0$, repeated $2^{N}-2^{N-2}$ times, and $4$, repeated $2^{N-2}$ times.
\item The $c_i$ and $d_i$ matrices are singular and have trace $2^N$.
\end{enumerate}
Returning to~\eqref{equ.plwwsaf} and substituting for the matrix elements $S_{k_{r_k s_k } }^z$, we find, after some algebra,
\[
\begin{split}
H_{I_{rs} }  &=  - \frac{{NJh_z^2 }}{{4h^2 }}\delta _{rs}  + \frac{{Jh_z^2 }}{{2h^2 }}\delta _{rs} \sum\limits_{i = 1}^N {\sin^2 \left\{ {(r_i  - r_{i + 1} ){\pi  \mathord{\left/
 {\vphantom {\pi  2}} \right.
 \kern-\nulldelimiterspace} 2}} \right\}}\\
&\\
& \quad - (1 - \delta _{rs} )\frac{{h_z J\left( {h^2  - h_z^2 } \right)^{1/2} }}{{2h^2 }}P_{rs}  + (1 - \delta _{rs} )\frac{{J\left( {h^2  - h_z^2 } \right)}}{{4h^2 }}Q_{rs}\,,
\end{split}
\]
where, (for $r\ne s$),
\[
\begin{split}
P_{rs}  &= \sum\limits_{i = 1}^N { c_{i_{rs} }\cos \left\{ {\left( {r_{i - 1}  + r_{i + 1} } \right)\frac{\pi }{2}} \right\} }\\
&\\
&= \delta _{\beta _{rs} ,N - 2}\,\cos \left\{ {\left( {r_{k - 1}  + r_{k + 1} } \right)\frac{\pi }{2}} \right\}
\end{split}
\]
and
\[
Q_{rs}  = \sum\limits_{i = 1}^N {\left( {2c_{i_{rs} }  - d_{i_{rs} } } \right)}  =2c_{rs}-d_{rs}=  - \delta _{\beta _{rs} ,N - 3}\,,
\]
where
\[
k = \sum\limits_{j = 1}^N {j\left( {r_j  - s_j } \right)^2 }  = \sum\limits_{j = 1}^N {j\left( {1 - \delta _{r_j s_j } } \right)}  = \sum\limits_{j = 1}^N {j\sin^2 \left\{ {\left( {r_j  - s_j } \right)\frac{\pi }{2}} \right\}}\,. 
\]

Explicitly,
\[
P_{rs}  = \left\{ {\begin{array}{*{20}c}
   1 & \mbox{ if }{\beta _{rs}  = N - 2\mbox{ and }r_{k - 1}  = 0 = r_{k + 1} }  \\
   &\\
   0 & \mbox{ if }{\beta _{rs}  < N - 2\mbox{ or }r_{k - 1}  + r_{k + 1}  = 1}  \\
   &\\
   { - 1} &\mbox{ if } {r_{k - 1}  = 1 = r_{k + 1} }  \\
\end{array}} \right.
\]
and
\[
Q_{rs}  = \left\{ {\begin{array}{*{20}c}
   { - 1} & \mbox{ if }{\beta _{rs}  = N - 3}  \\
   &\\
   0 & \mbox{ if }{\beta _{rs}  < N - 3}  \\
\end{array}} \right.\,.
\]
Putting the results together we finally have the matrix elements for the Ising interaction Hamiltonian, $H_I$, to be explicitly given by
\[
\begin{split}
H_{I_{rs}}  &=  - \frac{{NJ}}{4}\frac{{h_z^2 }}{{h^2 }}\delta _{rs}  + \frac{{Jh_z^2 }}{{2h^2 }}\delta _{rs} \sum\limits_{i = 1}^N {\sin^2 \left\{ {\left( {r_i  - r_{i + 1} } \right)\frac{\pi }{2}} \right\}}\\
&\\
&\quad - (1 - \delta _{rs} )\frac{{Jh_z \left( {h^2  - h_z^2 } \right)^{1/2} }}{{2h^2 }}\delta _{\beta _{rs} ,N - 2} \cos \left\{ {\left( {r_{k - 1}  + r_{k + 1} } \right)\frac{\pi }{2}} \right\}\\
&\\
&\qquad- (1 - \delta _{rs} )\frac{{J\left( {h^2  - h_z^2 } \right)}}{{4h^2 }}\delta _{\beta _{rs} ,N - 3}\,,
\end{split}
\]
where
\[
k = \sum\limits_{j = 1}^N {j\sin^2 \left\{ {\left( {r_j  - s_j } \right)\frac{\pi }{2}} \right\}}\,. 
\]
Since $H_{rs}  = H_{F_{rs} }  + H_{I_{rs} } $ we therefore have that the matrix elements of the Ising model in mutually orthogonal external magnetic fields are given by
\[
\begin{split}
H_{rs}&=h\delta_{rs}\sum_{i=1}^Nr_i-\frac{Nh}{2}\delta_{rs}  - \frac{{NJ}}{4}\frac{{h_z^2 }}{{h^2 }}\delta _{rs}  + \frac{{Jh_z^2 }}{{2h^2 }}\delta _{rs} \sum\limits_{i = 1}^N {\sin^2 \left\{ {\left( {r_i  - r_{i + 1} } \right)\frac{\pi }{2}} \right\}}\\
&\\
&\quad - (1 - \delta _{rs} )\frac{{Jh_z \left( {h^2  - h_z^2 } \right)^{1/2} }}{{2h^2 }}\delta _{\beta _{rs} ,N - 2} \cos \left\{ {\left( {r_{k - 1}  + r_{k + 1} } \right)\frac{\pi }{2}} \right\}\\
&\\
&\qquad- (1 - \delta _{rs} )\frac{{J\left( {h^2  - h_z^2 } \right)}}{{4h^2 }}\delta _{\beta _{rs} ,N - 3}\,,
\end{split}
\]

with $k$ as defined above.

\bigskip

Defining
\[
f = \frac{{h_z }}{h},\quad g = \frac{{(h^2  - h_z^2 )^{1/2} }}{h},\quad f^2+g^2=1\,,
\]
we have
\begin{equation}\label{equ.wgyg64p}
\begin{split}
H_{I _{rs}}  &=  - \frac{{NJf^2 }}{4}\delta _{rs}  + \frac{{Jf^2 }}{2}\delta _{rs} \sum\limits_{i = 1}^N {\sin ^2 \left\{ {\left( {r_i  - r_{i + 1} } \right)\frac{\pi }{2}} \right\}}\\
&\\
&\quad- (1 - \delta _{rs} )\delta _{\beta _{rs} ,N - 2} \frac{{Jfg}}{2}\cos \left\{ {\left( {r_{k - 1}  + r_{k + 1} } \right)\frac{\pi }{2}} \right\}\\
&\\
&\qquad- (1 - \delta _{rs} )\frac{{Jg^2 }}{4}\delta _{\beta _{rs} ,N - 3}
\end{split}
\end{equation}
and
\begin{equation}
\begin{split}
H_{rs}  &= m_r h\delta _{rs}  - \frac{{Nh}}{2}\delta _{rs}  - \frac{{NJf^2 }}{4}\delta _{rs}\\
&\\
&\quad + \frac{{Jf^2 }}{2}\delta _{rs} \sum\limits_{i = 1}^N {\sin ^2 \left\{ {\left( {r_i  - r_{i + 1} } \right)\frac{\pi }{2}} \right\}}\\
&\\
&\qquad- (1 - \delta _{rs} )\delta _{\beta _{rs} ,N - 2} \frac{{Jfg}}{2}\cos \left\{ {\left( {r_{k - 1}  + r_{k + 1} } \right)\frac{\pi }{2}} \right\}\\
&\\
&\qquad\quad- (1 - \delta _{rs} )\frac{{Jg^2 }}{4}\delta _{\beta _{rs} ,N - 3}\,,
\end{split}
\end{equation}
where
\[
m_r  = \sum\limits_{j = 1}^N {r_j } ,\quad k = \sum\limits_{j = 1}^N {j\sin ^2 \left\{ {\left( {r_j  - s_j } \right)\frac{\pi }{2}} \right\}}\,. 
\]

\section{Example application: ground state energy of weakly interacting spin~$1/2$ particles in external magnetic fields}
When the exchange integral $J$ is small, the Ising interaction term $H_I$ can be treated as a perturbation of $H_F$. In this section, we employ~\eqref{equ.wgyg64p} to find corrections, up to the fourth order in $J$, to the energy of the ground state of weakly interacting spin~$1/2$ particles in mutually orthogonal external magnetic fields. Since the ground state of $H_F$, the unperturbed system, is non-degenerate, we will apply the non-degenerate Rayleigh-Schr\"odinger perturbation theory.  

\bigskip

The following particular cases of~\eqref{equ.wgyg64p} will often be useful.
\begin{equation}\label{equ.dpaas1h}
H_{I_{ss} }  =  - \frac{{NJf^2 }}{4} + \frac{{Jf^2 }}{2}\sum\limits_{i = 1}^N {\sin ^2 \left\{ {\left( {s_i  - s_{i + 1} } \right)\frac{\pi }{2}} \right\}}\,.
\end{equation} 
In particular,
\begin{equation}\label{equ.yu6sixf}
H_{I_{00} }  =  - \frac{Nf^2}{4}J\,.
\end{equation}
For $s\ne t$
\begin{equation}\label{equ.c21w87f}
H_{I_{st} }  =  - \frac{{fgJ}}{2}\delta _{\beta _{st} ,N - 2} \cos \left\{ {\left( {s_{k - 1}  + s_{k + 1} } \right)\frac{\pi }{2}} \right\} - \frac{{g^2J }}{4}\delta _{\beta _{st} ,N - 3}\,,
\end{equation}
where
\[
k = \sum\limits_{j = 1}^N {j\sin ^2 \left\{ {\left( {r_j  - s_j } \right)\frac{\pi }{2}} \right\}}\,.
\]
In particular,
\begin{equation}\label{equ.bf9zq5p}
H_{I_{0t} }  =  - \frac{{fgJ}}{2}\delta _{\beta _{0t} ,N - 2}  - \frac{g^2J}{4}\delta _{\beta _{0t} ,N - 3}\,.
\end{equation}
Note also from~\eqref{equ.cvuwlu9} that
\begin{equation}\label{equ.c0vsjbp}
E_r  - E_s  =E_{rs}= (m_r  - m_s )h,\quad E_{0s}  =  - m_s h\,.
\end{equation}

\subsection{First order correction to the energy}

The first order correction to the energy of the ground state of $H_F$ is the expectation value of the perturbation $H_I$ in the ground state $\left| E_0 \right\rangle$ of $H_F$. 

\bigskip

Thus, quoting~\eqref{equ.yu6sixf}, we have
\begin{equation}\label{equ.dt1y3xp}
E_0^{(1)}  = \left\langle {H_I } \right\rangle _{\left| {E_0 } \right\rangle }  = \left\langle {E_0 } \right|H_I\left| {E_0 } \right\rangle  = H_{I_{00} }  =- \frac{{Nf^2}}{4}J\,.
\end{equation}

\subsection{Second order correction to the energy}
The second order correction to the energy of the ground state of $H_F$ is given by
\[
\begin{split}
E_0^{(2)}  &= \sum\limits_{s = 1}^{2^N  - 1} {\frac{{\left\langle {E_0 } \right|H_I \left| {E_s } \right\rangle \left\langle {E_s } \right|H_I \left| {E_0 } \right\rangle }}{{E_0  - E_s }}}\\
&\\
&= \sum\limits_{s = 1}^{2^N  - 1} {\frac{{\left| {H_{I_{0s} } } \right|^2 }}{E_{0s} }}\,.
\end{split}
\]

According to~\eqref{equ.bf9zq5p},
\[
H_{I_{0s} }  =  - \frac{{fgJ }}{2}\delta _{\beta _{0s} ,\,N - 2}  - \frac{g^2J}{4}\delta _{\beta _{0s} ,\,N - 3}\,. 
\]
We therefore see that contributions to $E_0^{(2)}$ come only from states with either $m_s=\sum s_i=1$ (corresponding to $\beta_{0s}=N-2$) or $m_s=\sum s_i=2$ (corresponding to $\beta_{0s}=N-3$ in the case when the two $\left| {\varepsilon_1 } \right\rangle$ states of the direct product state $\left| {E_s } \right\rangle$ are consecutive). A typical state with $m_s=1$ is the state 
\[
\left| {E_{2^{N  - 1}} } \right\rangle  = \left| {\varepsilon _1 } \right\rangle \left| {\varepsilon _0 } \right\rangle \left| {\varepsilon _0 } \right\rangle  \cdots \left| {\varepsilon _0 } \right\rangle  \cdots \left| {\varepsilon _0 } \right\rangle  \equiv (1,0,0, \cdots ,0, \cdots ,0)
\]
while a particular state with $m_s=2$ (and $\beta_{0s}=N-3$) is the state
\[
\left| {E_{_{3 \times 2^{N - 2} } } } \right\rangle  = \left| {\varepsilon _1 } \right\rangle \left| {\varepsilon _1 } \right\rangle \left| {\varepsilon _0 } \right\rangle  \cdots \left| {\varepsilon _0 } \right\rangle  \cdots \left| {\varepsilon _0 } \right\rangle  \equiv (1,1,0, \cdots ,0, \cdots ,0).
\]
Therefore
\[
H_{I_{0,2^{N - 1} } }  =  - \frac{fgJ}{2}\mbox{ and } H_{I_{0,3 \times 2^{N - 2} } }  =  - \frac{g^2J}{4}\,,
\]
and since there are $N$ vectors with $\beta_{0s}=N-2$ and N vectors with $\beta_{0s}=N-3$, and using~\eqref{equ.c0vsjbp}, we obtain
\begin{equation}\label{equ.okvffc1}
\begin{split}
E_0^{(2)}  &=  - \frac{{N\left| {H_{I_{0,2^{N - 1} } } } \right|^2 }}{h} - \frac{{N\left| {H_{I_{0,3 \times 2^{N - 2} } } } \right|^2 }}{{2h}}\\
&\\
&=  - \frac{{Nf^2 g^2 }}{{4h}}J^2 - \frac{{Ng^4 }}{{32h}}J^2\,.
\end{split}
\end{equation}

The results~\eqref{equ.dt1y3xp} and \eqref{equ.okvffc1} were also obtained in~\cite{adegoken}.

\subsection{Third order correction to the energy}
The third order correction to the energy of the ground state of $H_F$ is obtainable from the formula
\[
\begin{split}
E_0^{(3)}  &= \sum\limits_{s = 1}^{2^N  - 1} {\sum\limits_{t = 1}^{2^N  - 1} {\frac{{H_{I_{0s} } H_{I_{st} } H_{I_{t0} } }}{{E_{0s} E_{0t} }}} }  - H_{I_{00} } \sum\limits_{s = 1}^{2^N  - 1} {\frac{{\left| {H_{I_{0s} } } \right|^2 }}{{E_{0s}^2 }}}\\
&\\
&= \sum\limits_{s = 1}^{2^N  - 1} {\frac{{\left| {H_{I_{0s} } } \right|^2 H_{I_{ss} } }}{{E_{0s}^2 }}}  + 2\sum\limits_{s = 1}^{2^N  - 2} \, {\sum\limits_{t = s + 1}^{2^N  - 1} {\frac{{H_{I_{0s} } H_{I_{st} } H_{I_{t0} } }}{{E_{0s} E_{0t} }}} }  - H_{I_{00} } \sum\limits_{s = 1}^{2^N  - 1} {\frac{{\left| {H_{I_{0s} } } \right|^2 }}{{E_{0s}^2 }}}\\
&\\
&=S_1+S_2+S_3\,,
\end{split}
\]
where
\[
\begin{split}
&S_1=\sum\limits_{s = 1}^{2^N  - 1} {\frac{{\left| {H_{I_{0s} } } \right|^2 H_{I_{ss} } }}{{E_{0s}^2 }}}\,,\quad S_2=2\sum\limits_{s = 1}^{2^N  - 2}\,{\sum\limits_{t = s + 1}^{2^N  - 1} {\frac{{H_{I_{0s} } H_{I_{st} } H_{I_{t0} } }}{{E_{0s} E_{0t} }}} }\,,\\
&\\
&S_3=- H_{I_{00} } \sum\limits_{s = 1}^{2^N  - 1} {\frac{{\left| {H_{I_{0s} } } \right|^2 }}{{E_{0s}^2 }}}\,.
\end{split}
\]
Note that in the above derivation we made use of the following summation identity
\[
\sum\limits_{s = a}^M {\sum\limits_{t = a}^M {f_{st} } }  = \sum\limits_{s = a}^M {f_{ss} }  + \sum\limits_{s = a}^{M - 1} {\sum\limits_{t = s + 1}^M {\left( {f_{st}  + f_{ts} } \right)} }\,. 
\]
\subsubsection*{Evaluation of $S_1$}

\begin{itemize}
\item Contribution from states with $m_s=1$ ($ \Rightarrow \beta_{0s}=N-2$ ) 
\[
H_{I_{0s} }  =  - \frac{{fgJ}}{2}\;\mbox{ (from~\eqref{equ.bf9zq5p})},\quad H_{I_{ss} }  =  - \frac{{Nf^2 J}}{4} + f^2 J\;\mbox{ (from~\eqref{equ.dpaas1h})}
\]
The contribution of the $N$  states with $m_s=1$ to the sum $S_1$ is therefore
\[
{{N\frac{{f^2 g^2 J^2 }}{4}\left( { - \frac{{Nf^2 J}}{4} + f^2 J} \right)} \mathord{\left/
 {\vphantom {{N\frac{{f^2 g^2 J^2 }}{4}\left( { - \frac{{Nf^2 J}}{4} + f^2 J} \right)} {h^2 }}} \right.
 \kern-\nulldelimiterspace} {h^2 }}\,.
\]

\item Contribution from states with $m_s=2$ (provided that $\beta_{0s}=N-3$ ) 
\[
H_{I_{0s} }  =  - \frac{{g^2J}}{4},\quad H_{I_{ss} }  =  - \frac{{Nf^2 J}}{4} + f^2 J
\]

The $N$ states with $m_s=2$, $\beta_{0s}=N-3$ therefore contribute
\[
{{N\frac{{g^4 J^2 }}{{16}}\left( { - \frac{{Nf^2 J}}{4} + f^2 J} \right)} \mathord{\left/
 {\vphantom {{N\frac{{g^4 J^2 }}{{16}}\left( { - \frac{{Nf^2 J}}{4} + f^2 J} \right)} {\left( {4h^2 } \right)}}} \right.
 \kern-\nulldelimiterspace} {\left( {4h^2 } \right)}}
\]

to $S_1$.

\bigskip

Putting these results together we have

\begin{equation}\label{equ.hdnb3zl}
\begin{split}
S_1&={{\frac{{Nf^2 g^2 J^2 }}{4}\left( { - \frac{{Nf^2 J}}{4} + f^2 J} \right)} \mathord{\left/
 {\vphantom {{\frac{{Nf^2 g^2 J^2 }}{4}\left( { - \frac{{Nf^2 J}}{4} + f^2 J} \right)} {h^2 }}} \right.
 \kern-\nulldelimiterspace} {h^2 }}\\
&\\
&\qquad +{{\frac{{Ng^4 J^2 }}{{16}}\left( { - \frac{{Nf^2 J}}{4} + f^2 J} \right)} \mathord{\left/
 {\vphantom {{\frac{{Ng^4 J^2 }}{{16}}\left( { - \frac{{Nf^2 J}}{4} + f^2 J} \right)} {\left( {4h^2 } \right)}}} \right.
 \kern-\nulldelimiterspace} {\left( {4h^2 } \right)}}\,.
\end{split}
\end{equation}
\end{itemize}

\subsubsection*{Evaluation of $S_2$}
\[
S_2=2\sum\limits_{s = 1}^{2^N  - 2}\,{\sum\limits_{t = s + 1}^{2^N  - 1} {\frac{{H_{I_{0s} } H_{I_{st} } H_{I_{t0} } }}{{E_{0s} E_{0t} }}} }\,. 
\]
In each term of the sum, one of four different scenarios is possible, namely, \mbox{$m_s=1=m_t$} or \mbox{$m_s=2=m_t$} or \mbox{$m_s=1, m_t=2$} or \mbox{$m_s=2,\,m_t=1$}. We look at each possible situation in turn.
\begin{itemize} 
\item Contribution to $S_2$ when $m_s=1=m_t$

In this case, for each $s$~vector, there are two possible $t$~vectors for which the matrix element $H_{I_{st}}$ does not vanish, as typified below:
\[
\left. \begin{array}{l}
 s:(0,1,0,0, \cdots ,0,0)\qquad s:(0,1,0,0, \cdots ,0,0) \\ 
 t:(1,0,0,0, \cdots ,0,0)\qquad\, t:(0,0,1,0, \cdots ,0,0) \\ 
 \end{array} \right\}\,\beta _{st}  = N - 3\,.
\]
In such a situation,
\[
H_{I_{st}}=-\frac{g^2J}{4}\,.
\]
We also have
\[
\begin{split}
&H_{I_{0s} }  =  - \frac{{fgJ}}{2}(s \ne 0,\beta _{0s}  = N - 2)\\
&\mbox{and}\\
&H_{I_{t0} }  = H_{I_{0t} }  =  - \frac{{fgJ}}{2}(t \ne 0,\beta _{0t}  = N - 2)\,.
\end{split}
\]
Since there are $N$ $m_s=1$ states, the contribution to the sum $S_2$ when $m_s=1=m_t$ is
\[
\begin{split}
&{{\left( {2N \cdot 2 \cdot {{ - fgJ} \mathord{\left/
 {\vphantom {{ - fgJ} 2}} \right.
 \kern-\nulldelimiterspace} 2} \cdot  - {{g^2 J} \mathord{\left/
 {\vphantom {{g^2 J} 4}} \right.
 \kern-\nulldelimiterspace} 4} \cdot {{ - fgJ} \mathord{\left/
 {\vphantom {{ - fgJ} 2}} \right.
 \kern-\nulldelimiterspace} 2}} \right)} \mathord{\left/
 {\vphantom {{\left( {2N \cdot 2 \cdot {{ - fgJ} \mathord{\left/
 {\vphantom {{ - fgJ} 2}} \right.
 \kern-\nulldelimiterspace} 2} \cdot  - {{g^2 J} \mathord{\left/
 {\vphantom {{g^2 J} 4}} \right.
 \kern-\nulldelimiterspace} 4} \cdot {{ - fgJ} \mathord{\left/
 {\vphantom {{ - fgJ} 2}} \right.
 \kern-\nulldelimiterspace} 2}} \right)} {\left( { - h \cdot  - 2h} \right)}}} \right.
 \kern-\nulldelimiterspace} {\left( { - h \cdot  - 2h} \right)}}\\
 &=  - \frac{{Nf^2 g^4 J^3 }}{{8h^2 }}
\end{split}
\]
\item Contribution to $S_2$ when $m_s=2=m_t$

As in the previous case, for each $s$~vector, there are only two possible $t$~vectors for which the matrix element $H_{I_{st}}$ does not vanish, as typified below:
\[
\left. \begin{array}{l}
 s:(1,1,0,0, \cdots ,0,0)\qquad s:(1,1,0,0, \cdots ,0,0) \\ 
 t:(1,0,1,0, \cdots ,0,0)\qquad\, t:(0,1,0,0, \cdots ,0,1) \\ 
 \end{array} \right\}\,\beta _{st}  = N - 3\,.
\]
In such a situation,
\[
H_{I_{st}}=-\frac{g^2J}{4}\quad\mbox{but }H_{I_{t0}}=H_{I_{0t}}=0\mbox{ since $\beta_{t0}=N-4$}\,.
\]
There is therefore zero contribution to $S_2$ when $m_s=2=m_t$.

\item Contribution to $S_2$ when $m_s=2,\,m_t=1$

In this case, typical situations with an $s$~vector and the two $t$~vectors for which $H_{I_{rs}}$ does not vanish are depicted below
\[
\left. \begin{array}{l}
 s:(1,1,0,0, \cdots ,0,0)\qquad s:(1,1,0,0, \cdots ,0,0) \\ 
 t:(0,1,0,0, \cdots ,0,0)\qquad\, t:(1,0,0,0, \cdots ,0,0) \\ 
 \end{array} \right\}\,\beta _{st}  = N - 2\,.
\]
From~\eqref{equ.c21w87f} we have
\[
H_{I_{st} }  =  - \frac{{fgJ}}{2}\cos({\pi/2})=0\,,
\]
signifying a zero contribution to the $S_2$ sum.

\item Contribution to $S_2$ when $m_s=1,\,m_t=0$

Here as in the previous case we have \mbox{$H_{I_{st} }  =  - fgJ/2\cos({\pi/2})=0$}, so that again there is zero contribution to the $S_2$ sum. 

Adding all the contributions we have
\begin{equation}\label{equ.aw0zn0l}
S_2=  - \frac{{Nf^2 g^4 J^3 }}{{8h^2 }}\,.
\end{equation}

\subsubsection*{Evaluation of $S_3$}
\[
S_3=- H_{I_{00} } \sum\limits_{s = 1}^{2^N  - 1} {\frac{{\left| {H_{I_{0s} } } \right|^2 }}{{E_{0s}^2 }}}
\]
From~\eqref{equ.yu6sixf}, \eqref{equ.bf9zq5p} and \eqref{equ.c0vsjbp} we have immediately that
\begin{equation}\label{equ.bwwwwx9}
S_3  = \frac{{Nf^2 J}}{4}\left( {{{\frac{{Nf^2 g^2 J^2 }}{4}} \mathord{\left/
 {\vphantom {{\frac{{Nf^2 g^2 J^2 }}{4}} {h^2 }}} \right.
 \kern-\nulldelimiterspace} {h^2 }} + {{\frac{{Ng^4 J^2 }}{{16}}} \mathord{\left/
 {\vphantom {{\frac{{Ng^4 J^2 }}{{16}}} {\left( {4h^2 } \right)}}} \right.
 \kern-\nulldelimiterspace} {\left( {4h^2 } \right)}}} \right)\,.
\end{equation}

\end{itemize}

Finally combining~\eqref{equ.hdnb3zl}, \eqref{equ.aw0zn0l} and~\eqref{equ.bwwwwx9}, we obtain the third order correction to the energy of the ground state of $H_F$ as
\begin{equation}\label{equ.ytd8b9l}
E_0^{(3)}  =  - \frac{{7Nf^2 g^4 J^3 }}{{64h^2 }} + \frac{{Nf^4 g^2 J^3 }}{{4h^2 }}\,.
\end{equation}

\subsection{Fourth order correction to the energy}

The fourth order correction to the energy of the ground state of $H_F$ is given by the standard \mbox{Rayleigh-Schr\"odinger} perturbation formula
\[
\begin{split}
E_0^{(4)} &= \sum\limits_{s = 1}^{2^N  - 1} {\sum\limits_{t = 1}^{2^N  - 1} {\sum\limits_{u = 1}^{2^N  - 1} {\frac{{H_{I_{0s} } H_{I_{st} } H_{I_{tu} } H_{I_{u0} } }}{{E_{0s} E_{0t} E_{0u} }}} } }  - H_{I_{00} } \sum\limits_{s = 1}^{2^N  - 1} {\sum\limits_{t = 1}^{2^N  - 1} {\frac{{H_{I_{0s} } H_{I_{st} } H_{I_{t0} } }}{{E_{0s} E_{0t}^2 }}} }\\
&\\
&\qquad- H_{I_{00} } \sum\limits_{s = 1}^{2^N  - 1} {\sum\limits_{t = 1}^{2^N  - 1} {\frac{{H_{I_{0s} } H_{I_{st} } H_{I_{t0} } }}{{E_{0s}^2 E_{0t} }}} }  + H_{I_{00} }^2 \sum\limits_{s = 1}^{2^N  - 1} {\frac{{\left| {H_{I_{0s} } } \right|^2 }}{{E_{0s}^3 }}}  - E_0^{(2)} \sum\limits_{s = 1}^{2^N  - 1} {\frac{{\left| {H_{I_{0s} } } \right|^2 }}{{E_{0s}^2 }}}\,.
\end{split}
\]
 
Calculations completely analogous to those in the previous sections, but much more involved, give $E_0^{(4)}$ as
\begin{equation}\label{equ.xy15rp6}
E_0^{(4)}  = -\frac{{13Nf^2 g^6 }}{{192h^3 }}J^4  + \frac{{55Nf^4 g^4 }}{{128h^3 }}J^4  - \frac{{Nf^6 g^2 }}{{4h^3 }}J^4  - \frac{{Ng^8 }}{{2048h^3 }}J^4\,.
\end{equation}
\subsection{Approximate analytical expression for the ground state energy per spin for weakly interacting spin~$1/2$ particles in external magnetic fields}
Adding the energy corrections~\eqref{equ.dt1y3xp}, \eqref{equ.okvffc1}, \eqref{equ.ytd8b9l} and \eqref{equ.xy15rp6} to the ground state energy \mbox{(obtained by setting $m_r=0$ in~\eqref{equ.cvuwlu9})} of the non-interacting spin~$1/2$ particles in external magnetic fields we therefore find, to the fourth order in the exhange integral, $J$, that the energy of the ground state, $E_{0_{IF}}$, of the one dimensional Ising model in mutually orthogonal external magnetic fields, for $N$ spin sites is given by
\[
\begin{split}
E_{0_{IF} }  &\approx  - \frac{{Nh}}{2} - \frac{{Nf^2 }}{4}J - \frac{{Nf^2 g^2 }}{{4h}}J^2  - \frac{{Ng^4 }}{{32h}}J^2  - \frac{{7Nf^2 g^4 }}{{64h^2 }}J^3  + \frac{{Nf^4 g^2 }}{{4h^2 }}J^3\\
&\\
&\qquad- \frac{{13Nf^2 g^6 }}{{192h^3 }}J^4  + \frac{{55Nf^4 g^4 }}{{128h^3 }}J^4  - \frac{{Nf^6 g^2 }}{{4h^3 }}J^4  - \frac{{Ng^8 }}{{2048h^3 }}J^4\,,
\end{split}
\]
that is
\[
\begin{split}
\frac{{e_0 }}{{\varepsilon _0 }} &\approx 1 + \frac{{f^2 }}{4}z + \left( {\frac{{g^2 }}{{64}} + \frac{{f^2 }}{8}} \right)g^2 z^2  + \left( {\frac{{7f^2 g^2 }}{{256}} - \frac{{f^4 }}{{16}}} \right)g^2 z^3\\
&\\
&\qquad\qquad+ \left( {\frac{{g^6 }}{{16384}} + \frac{{13f^2 g^4 }}{{1536}} - \frac{{55f^4 g^2 }}{{1024}} + \frac{{f^6 }}{{32}}} \right)g^2 z^4\,,
\end{split}
\]
where $e_0=E_{0_{IF} }/N$ is the ground state energy per spin, $\varepsilon_0=-h/2$ and \mbox{$z=-J/\varepsilon_0$}.

\bigskip

Since \mbox{$f^2+g^2=1$}, we can also write
\[
\begin{split}
\frac{{e_0 }}{{\varepsilon _0 }} &\approx 1 + \left( {\frac{1}{4} - \frac{{g^2 }}{4}} \right)z + \left( {\frac{1}{8} - \frac{7}{{64}}g^2 } \right)g^2 z^2\\
&\\
&\qquad + \left( { - \frac{1}{{16}} + \frac{{39}}{{256}}g^2  - \frac{{23}}{{256}}g^4 } \right)g^2 z^3\\
&\\
&\qquad\quad+ \left( {\frac{1}{{32}} - \frac{{151}}{{1024}}g^2  + \frac{{161}}{{768}}g^4  - \frac{{4589}}{{49152}}g^6 } \right)g^2 z^4\,,
\end{split}
\]
or, in a more compact form,
\begin{equation}\label{equ.skppgig}
\frac{{e_0 }}{{\varepsilon _0 }} \approx 1 + \frac{{f^2 }}{4}z + \sum\limits_{m = 2}^4 {\left\{ {z^m \sum\limits_{k = 0}^{m - 1} {( - 1)^{m - k} c_k^{(m)} (g^2 )^{k + 1} } } \right\}}\,,
\end{equation}
with
\[
\begin{split}
\quad c_0^{(m)}  = \frac{{( - 1)^m }}{{2^{m + 1} }},\quad &m = 1,2, \ldots\,,\\
&\\
& c_1^{(2)}  = \frac{7}{{64}},\\
&\\
& c_1^{(3)}  = \frac{{39}}{{256}},\, c_2^{(3)}  = \frac{{23}}{{256}}\\
&\\
& c_1^{(4)}  = \frac{{151}}{{1024}},\, c_2^{(4)}  = \frac{{161}}{{768}},\, c_3^{(4)}  = \frac{{4589}}{{49152}}\,.
\end{split}
\]
Note that when $f=0$, then 
\[
\frac{e_0}{\varepsilon_0}\approx 1+{\frac {1}{64}}{z}^{2}+{\frac {1}{16384}}{z}^{4}\,,
\]
in perfect agreement with the exact result for the ground state energy of the transverse field Ising model~\cite{pfeuty}:
\[
\frac{e_0}{\varepsilon_0}=\frac{(4+z)}{2\pi}\mathcal{E}\left[\frac{4\sqrt{z}}{4+z}\right] =1+{\frac {1}{64}}{z}^{2}+{\frac {1}{16384}}{z}^{4}+O \left( {z}^{6} \right)\,,
\]
where $\mathcal{E}$ is a complete elliptic integral of the second kind.

\bigskip

The form of~\eqref{equ.skppgig} suggests an exact result for the ground state energy per spin of the Ising model in external magnetic fields:
\[
\frac{{e_0 }}{{\varepsilon _0 }} = 1 + \frac{{f^2 }}{4}z + \sum\limits_{m = 2}^\infty {\left\{ {z^m \sum\limits_{k = 0}^{m - 1} {( - 1)^{m - k} c_k^{(m)} (g^2 )^{k + 1} } } \right\}}\,, 
\]
where $c_k^{(m)}$ are positive rational numbers, and in particular, \mbox{$c_0^{(m)}=(-1)^m/2^{(m+1)}$} for~$m\ge 2$.

\subsection{Estimation of various order parameters for the Ising model in mutually orthogonal external magnetic fields}
The knowledge of $e_0$ allows the derivation of approximate analytic expressions for physical quantities such as the magnetization in each direction and the spin-spin correlation function for neighbouring spins. 
\subsubsection{Magnetization}
Invoking the \mbox{Hellmann-Feynman} rule in~\eqref{equ.ir12ul8} gives for the \mbox{$x-$magnetization}
\[
m_x  = \frac{2}{N}\left\langle {\sum\limits_{i = 1}^N {S_i^x } } \right\rangle _{\left| {E_{0_{IF} } } \right\rangle }  =  - 2\frac{{\partial e_0 }}{{\partial h_x }} =  - 2\frac{{\partial h}}{{\partial h_x }}\frac{{\partial e_0 }}{{\partial h_x }} =  - 2\frac{{h_x }}{h}\frac{{\partial e_0 }}{{\partial h_x }}
\]
and similar expressions for $m_y$ and $m_z$, the $y-$ and \mbox{$z-$magnetizations}. 

\bigskip

According to~\eqref{equ.skppgig},
\[
e_0  \approx  - \frac{h}{2} - \frac{h}{8}zf^2  - \frac{h}{2}\sum\limits_{m = 2}^4 {\left\{ {z^m \sum\limits_{k = 0}^{m - 1} {( - 1)^{m - k} c_k^{(m)} \left( {g^2 } \right)^{k + 1} } } \right\}}\,, 
\]
so that for $h\ne 0$ we obtain
\begin{equation}
\begin{split}
\frac{{\partial e_0 }}{{\partial h}} &\approx  - \frac{1}{2} + \frac{{zf^2 }}{4}\\
&\\
&\qquad+ \sum\limits_{m = 2}^4 {\left\{ {z^m \sum\limits_{k = 0}^{m - 1} {( - 1)^{m - k} c_k^{(m)} g^{2k} \left( { - (k + 1)f^2  + \frac{{(m - 1)}}{2}g^2 } \right)} } \right\}}\,.
\end{split}
\end{equation}
Thus for $h_z<h\ne 0$,
\[
m_z  \approx f - \frac{z}{2}f^3  + \sum\limits_{m = 2}^4 {\left\{ {z^m \sum\limits_{k = 0}^{m - 1} {( - 1)^{m - k} c_k^{(m)} g^{2k} \left( {2(k + 1)f^3  - (m - 1)g^2 f} \right)} } \right\}}\,, 
\]
and for $h_x<h\ne 0$ and $h_y<h\ne 0$, respectively,
\[
m_x  \approx \frac{h_x}{h} - \frac{z}{2}\frac{h_x}{h} f^2  + \frac{h_x}{h} \sum\limits_{m = 2}^4 {\left\{ {z^m \sum\limits_{k = 0}^{m - 1} {( - 1)^{m - k} c_k^{(m)} g^{2k} \left( {2(k + 1)f^2  - (m - 1)g^2} \right)} } \right\}} 
\]
and
\[
m_y  \approx \frac{h_x}{h} - \frac{z}{2}\frac{h_y}{h} f^2  + \frac{h_y}{h} \sum\limits_{m = 2}^4 {\left\{ {z^m \sum\limits_{k = 0}^{m - 1} {( - 1)^{m - k} c_k^{(m)} g^{2k} \left( {2(k + 1)f^2  - (m - 1)g^2} \right)} } \right\}}\,. 
\]
Note that in the absence of interaction, ($z=0,h\ne 0$), $m_x^2+m_y^2+m_z^2=1$.
\subsubsection{Nearest neighbour \mbox{spin-spin} correlation}
The \mbox{spin-spin} correlation, $c_{i,i + 1}$, is given by
\[
c_{i,i + 1}  = \frac{4}{N}\left\langle {\sum\limits_{i = 1}^N {S_i^z S_{i + 1}^z } } \right\rangle _{\left| {E_{0_{IF} } } \right\rangle }  =  - 4\frac{{\partial e_0 }}{{\partial J}} =  - 4\frac{{\partial z}}{{\partial J}}\frac{{\partial e_0 }}{{\partial z}} =  - \frac{8}{h}\frac{{\partial e_0 }}{{\partial z}}\,,
\]
yielding
\[
c_{i,i + 1}  = f^2  + 4\sum\limits_{m = 2}^4 {\left\{ {mz^{m - 1} \sum\limits_{k = 0}^{m - 1} {( - 1)^{m - k} c_k^{(m)} \left( {g^2 } \right)^{k + 1} } } \right\}}\,. 
\]
Note that in the absence of interaction, $z=0$, we have $c_{i,i + 1}  = f^2$ while $h=h_z$ gives $c_{i,i + 1}  = 1$.
\section{Conclusion}
We have given an explicit matrix representation for the Hamiltonian of the Ising model in mutually orthogonal external magnetic fields, with basis the eigenstates of a system of non-interacting spin~$1/2$ particles in external magnetic fields. We subsequently applied our results to obtain an analytical expression for the ground state energy per spin, to the fourth order in the exchange integral, for the Ising model in perpendicular external fields. Since the Hamiltonian of the non-interacting spin~$1/2$ particles in external magnetic fields is a Hermitian operator that lives in a $2^N$ dimensional Hilbert space, its eigenstates form a complete orthonormal basis, suitable for giving matrix representations for any operator living in the same Hilbert space and with the same conditions at the boundary.

\end{document}